\documentclass[preprint,12pt]{elsarticle}

\usepackage{amssymb}
\usepackage{xcolor}
\usepackage{cleveref}
\usepackage{todonotes}
\usepackage{caption}
\usepackage[utf8]{inputenc}

\journal{Current Opinion in Solid State and Materials Science}

\begin{document}

\begin{frontmatter}



\title{Spintronic devices as next-generation computation accelerators}


\author[GU]{Victor H. González}

\affiliation[GU]{organization={Department of Physics, University of Gothenburg},
            addressline={Kemivägen 9}, 
            city={Gothenburg},
            postcode={41296}, 
            state={Västra Götaland},
            country={Sweden}}
\author[GU]{Artem Litvinenko}
\author[GU,Tohoku1,Tohoku2]{Akash Kumar}
\author[GU]{Roman Khymyn}
\author[GU,Tohoku1,Tohoku2]{Johan {\AA}kerman}

\affiliation[Tohoku1]{organization={Research Institute of Electrical Communication (RIEC), Tohoku University},
            addressline={2-1-1 Katahira, Aoba-ku}, 
            city={Sendai},
            postcode={980-8577}, 
            state={Tohoku Region},
            country={Japan}}

\affiliation[Tohoku2]{organization={Center for Science and Innovation in Spintronics, Tohoku University},
            addressline={2-1-1 Katahira, Aoba-ku}, 
            city={Sendai},
            postcode={980-8577}, 
            state={Tohoku Region},
            country={Japan}}

\begin{abstract}
The ever increasing demand for computational power combined with the predicted plateau for the miniaturization of existing silicon-based technologies has made the search for low power alternatives an industrial and scientifically engaging problem. In this work, we explore spintronics-based Ising machines as hardware computation accelerators. We start by presenting the physical platforms on which this emerging field is being developed, the different control schemes and the type of algorithms and problems on which these machines outperform conventional computers. We then benchmark these technologies and provide an outlook for future developments and use-cases that can help them get a running start for integration into the next generation of computing devices.
\end{abstract}

\begin{graphicalabstract}
\includegraphics[width=\linewidth]{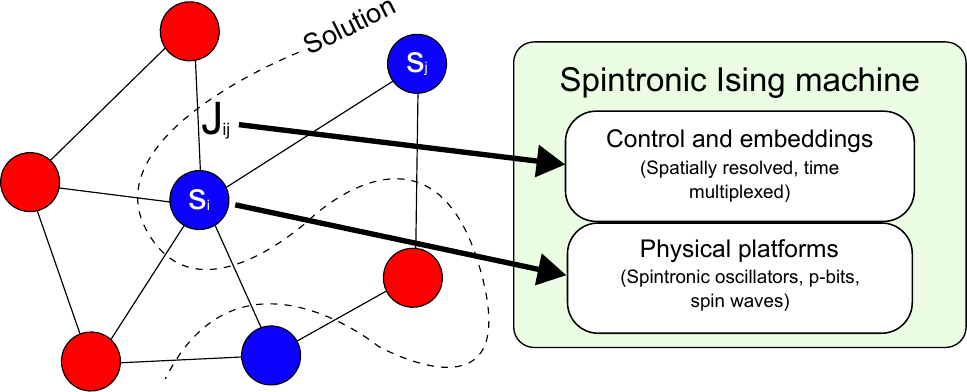}
\end{graphicalabstract}

\begin{highlights}
\item Spintronic-based computing architectures are a promising alternative in the emerging field of Ising machines due to their potential in parallelization and low power consumption.
\item An Ising machine condensed-matter-based annealer can be applied to the same use cases as a quantum annealer, with the advantage of operating at room temperature at a low operational cost.
\end{highlights}


\end{frontmatter}


\section{Introduction}
Non-Von Neumann (or unconventional) computation is an alternative to existing computation paradigms that aims to accelerate computing by enabling in-memory computing and large-scale parallelization. Interest in nano-magnetic materials and devices for unconventional computing has been growing in research and industrial circles due to the opportunities that they present to expand and complement existing computing architectures and enable controllable parallelization~\cite{grollier2020neuromorphic}. In this work, we present recent theoretical and experimental work on spintronic devices for combinatorial optimization using the Ising machine (IM), a type of unconventional computing architecture. We benchmark and compare the proposed schemes, and discuss possible future developments in the field. The growing interest in physical IM and IM-inspired algorithms is the evidence of the potential that this technology holds for acceleration in finding optimal solutions to exponentially growing NP-complete and NP-hard problems, such as graph partitioning~\cite{Ushijima2017-graph-partitioning-d-wave}, circuit layout design~\cite{Barahone1988-circuit-design}, graph colouring~\cite{Wang2021-OIM}, condition satisfiability~\cite{Sharma2023-augmented-IM}, traveling salesman~\cite{Sutton2017-optimization-nanomagnets}, and weighted knapsack~\cite{Ibarra1975-knapsack}.

At its core, an IM is an embedding of an Ising model into an ensemble of coupled physical systems such that their steady-state solutions allow us to minimize the Ising Hamiltonian: 
\begin{equation}
    \label{eq:hamiltonian}
    H (s_1,...,s_N) = -\sum_{i<j} J_{ij} s_i s_j + \sum_{i=1}^N h_i s_i 
\end{equation}
In the Ising model, $s_i=\pm 1$ are the binary spins, $J_{ij}$ is the coupling between the i-th and j-th spins and $h_i$ is the individual biasing field over the i-th spin.

As it has been shown it is possible to map many NP-complete and NP-hard problems to the Ising Hamiltonian~\cite{Lucas2014IsingFormulations}, progress into these embeddings has been fueled by the potential applications that a high-speed solver would have in both academia and industry. Examples of such applications include protein folding~\cite{babej2018-protein-folding-quantum-annealer}, molecular assembly~\cite{ichikawa2023-crystal-assembly}, logistic scheduling~\cite{venturelli2016-job-scheduling-quantum-annealer, Shimada2021-job-scheduling-Ising-machine} and econometric calculations~\cite{Ibarra1975-knapsack}. Because the solution spaces of these problems grow exponentially with the size of the problem, physics-based energy minimization of the Ising model is an appealing alternative to heuristic algorithms dependent on conventional computing architectures and sequential operations. 

Different ensembles have been proposed as possible IM hardware such as existing CMOS technology~\cite{tsukamoto2017accelerator, yamaoka2015-20k-spins}, electrical oscillators~\cite{Afoakwa-BRIM}, interfering light~\cite{Pierangeli2019-photonic-annealer}, optical pulses~\cite{McMahon2016Sci100CIM, Inagaki2016Sci2000CIM, Honjo2021SciAdv100kCIM, marandi2014networkCIM4delaylines}, memristor crossbar arrays~\cite{Bojnordi2016-memristor-IM, cai2020-memristor-hopfield-networks}, Josephson junctions~\cite{albash2018-d-wave-computer, boixo2016-multiqubit-tunnelling-quantum-annealers}, and even single atoms~\cite{Kiraly2021-atomic-boltzmann-machine}; with the latter two making use of the quantum adiabatic theorem to tunnel to the ground state. These so-called quantum annealers are useful for combinatorial optimization tasks, albeit at a high operational cost due to the precise thermal and temporal stability necessary to perform any computation. That is where spintronic devices come into play, as their high operational speeds, low fabrication cost, and room temperature stability make them ideal for tackling a range of problems that do not require the superposition or tunneling of quantum systems.

\begin{figure}
    \centering
    \includegraphics[width=\linewidth]{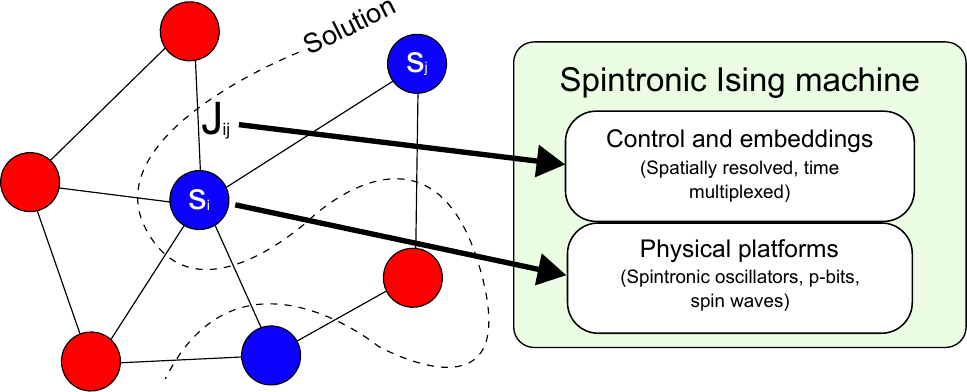}
    \caption{\textbf{Building blocks necessary for the successful development of spintronic computation accelerators} The graph associated to the Ising Hamiltonian of an NP problem has to be constructed on a physical platform and then controlled using various mechanisms. The energy minimization of the physical system provides us with the solution.}
    \label{fig:accelerator-building-blocks}
\end{figure}

In the following sections, we will present the latest efforts to build IMs using spintronic devices. We will do this by taking a bottom-up approach, as shown in fig \ref{fig:accelerator-building-blocks}. First, we will introduce the physical platforms on which the artificial spins are mapped, their principles of operation, and their features. As the layer that does the processing, the capabilities of actual devices impinge on the kind of problems that can realistically be solved. Building upon that foundation, we will discuss the construction and control of the coupled ensembles: different challenges and opportunities that each platform has to offer depending on the specific binarizing and coupling mechanisms. We will show the embeddings achieved by different authors and how the choice of transformation of the artificial spin phase space can better utilize the hardware's capabilities. We round up our presentation by benchmarking and discussing the devices, their applications, and the current trends on the subject followed by a brief outlook for this emerging field.

\section{Physical platforms}

\subsection{Spintronic oscillator-based Ising machines}

Spintronic oscillators are a category of devices that have received a lot of interest from the condensed matter and electronics communities~\cite{Chen2016procieee} as they present ample integration opportunities with existing CMOS technology~\cite{zahedinejad2018cmos} while employing spin phenomena to transduce direct electric currents into microwave frequency signals~\cite{slonczewski1996jmmm,berger1996prb,Kiselev2003, ralph2008jmmm} with fast and wideband frequency tunability and with relatively narrow linewidth. Due to the magnetodynamical origin of these signals, they can be modulated using different means such as light~\cite{Salomoni-all-optical-switching-mtjs, muralidhar2022optothermal}, current~\cite{Dussaux2010-vortex-mtjs, Zahedinejad2017IEEE}, electric~\cite{wang2012electric-field-mtjs, liu2017controlling-shnos-eletric-field,litvinenko2021analog} and magnetic fields~\cite{hai2009-electromotive-force-mtjs, giordano2014-shnos-magnetic-field}, magnetic anisotropies~\cite{nozaki2010-VCMA-MTJs, Zhang2017-vcma-stnos, Fulara2020-VCMA, Gonzalez2022-VCMA-SNHO, sidi2022size-MTJs} and temperature~\cite{prejbeanu2004-thermally-assisted-mtjs, Li2004-thermally-assisted-reversal}. Because of these features, they have attracted considerable interest in the spintronics community, as evidenced in their use on the already existing magneto-resistive random access memory (MRAM), and the potential future applications as microwave sources and detectors~\cite{bendjeddou2023electrical}, wireless communication transceivers and artificial spins for IMs.

\begin{figure}
    \centering
    \includegraphics[width=\linewidth]{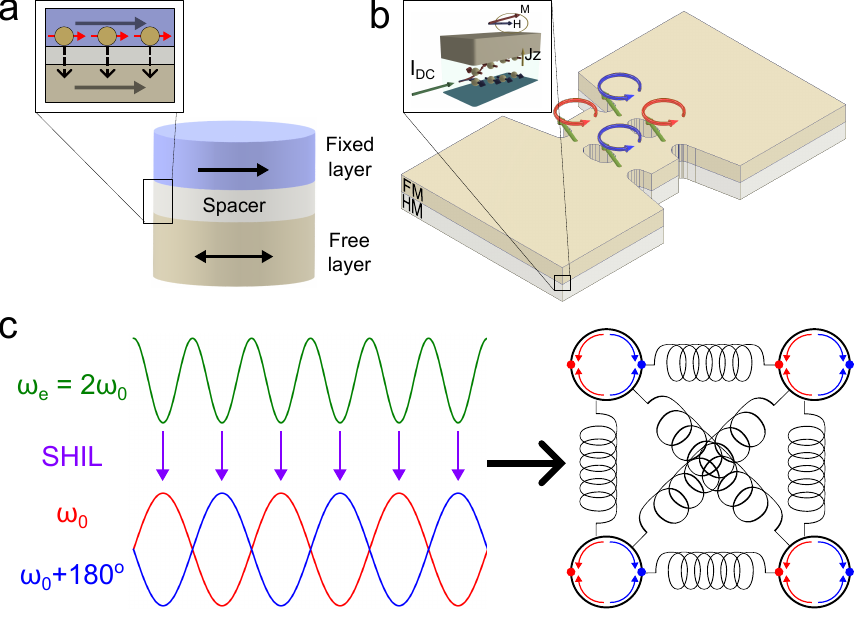}
    \caption{\textbf{Spintronic oscillators, their operation principles and second harmonic injection locking. a.} Diagram of a magnetic tunneling junction (MTJ), as current flows across the fixed layer it is spin polarized and it tunnels across the insulating spacer to generate precession in the free layer. \textbf{b.} Diagram and operation principle of an array of spin-Hall nano-oscillators (SHNOs). In SHNOs, the spin Hall effect depicted in the inset, generates a spin polarized current in the heavy metal (HM) which in turn generates precesion in the ferromagnetic (FM) layer. \textbf{c.} During second harmonic injection locking (SHIL), the oscillators are driven at twice their natural frequency and their phases binarize to $0^o$ and $180^o$ with respect to the reference signal. Networks of these binarized oscillators are used to construct Ising machines, as pictured in the diagram to the left.}
    \label{fig:spintronic-oscillators}
\end{figure}

Depending on the mechanism for the generation of the spin currents, spintronic oscillators fall into two categories: spin-torque nano-oscillators (STNOs) and spin Hall nano-oscillators (SHNO). Both employ the spin-transfer-torque (STT) from a spin current to compensate for the natural damping of ferromagnetic materials and induce precession in the magnetization of thin films. The difference between the two lies in both the generation mechanism for the spin current, and whether it is accompanied by a charge current (spin-polarized current \emph{vs.}~pure spin current), as shown in the insets of Fig.\ref{fig:spintronic-oscillators}a and b. In the case of STNOs, a spin-polarized current is generated when electrons traverse a magnetized ferromagnetic layer (fixed layer) due to spin splitting at the Fermi surface of the magnetic metal ~\cite{Dyakonov1971jetp}. On the other hand, SHNOs use the spin Hall effect (SHE)~\cite{Sinova2015} present in heavy metals with large spin-orbit coupling to produce a transverse pure spin current and inject them into an adjacent ferromagnetic layer~\cite{Liu2012prl,Demidov2012b}. 
 
The most popular type of STNOs is the magnetic tunnel junction (MTJ), thanks to its higher output power~\cite{Kiselev2003,Chen2016procieee, Houshang2018-nanocontact-mtjs}. When a current is injected through the fixed layer, it gets spin-polarized and it tunnels across the insulator (spacer) to the free layer, where it exerts STT. In an MTJ, the probability electrons tunneling between the fixed and free layers depends on the relative orientation of their magnetic moments. When the magnetic moments are parallel, the electrons can tunnel more easily and the resistance decreases. Conversely, in an anti-parallel alignment, the resistance increases. This allows MTJs to act as oscillators, but also ultra fast binary switching elements. The magneto-resistive feature coupled with their GHz operation frequency have made MTJs very appealing for data storage~\cite{Akerman2005sci}, magnetic sensors~\cite{egelhoff2009-picoTesla-sensors-mtjs}, and their switching dynamics make them ideal for stochastic elements in probabilistic computing paradigms~\cite{fukushima2014spin}.

In the SHNO space, the nanoconstriction architecture, shown in Fig. \ref{fig:spintronic-oscillators}b, remains the most popular~\cite{Demidov2014}. The nanoconstriction geometry allows for compensation of damping in the constricted region, inducing  localized microwave frequency precession of the magnetization~\cite{Demidov2014}. The nano-scale dimensions of SHNOs facilitate high-frequency operations and allow for significant miniaturization of the devices~\cite{durrenfeld2017nanoscale,behera2023ultra}. SHNOs are characterized by their wide frequency tunability~\cite{zahedinejad2018cmos, choi2022voltage, Fulara2020-VCMA}, nanoscale size~\cite{behera2023ultra, behera2022energy}, and strong non-linear behavior, which are highly desirable for unconventional computing applications due to their easily accessible free ferromagnetic layer~\cite{kumar2023book}. 


The collective dynamics of arrays of MTJs and SHNOs in both one-dimensional chains and two-dimensional arrays open up new possibilities in computing and signal processing~\cite{awad2017natphys,zahedinejad2020nt,kumar2023robust,litvinenko2023phase}. By coupling multiple oscillators, it is possible to achieve mutual synchronization, where the oscillators lock to a common frequency and phase. This is a major step towards the realization of oscillator-based computing paradigms, including neuromorphic computing and Ising machines. Regardless of origin, the precession modes generated allow one to treat the devices as GHz frequency oscillators and construct hardware solvers using oscillator-based Ising machines (OIMs).

In the OIM formalism~\cite{Wang2021-OIM}, second harmonic injection locking (SHIL) is used to binarize the phase dynamics of oscillators, and physical coupling between them allows one to map the Ising Hamiltonian, as shown in fig\ref{fig:spintronic-oscillators}c. In SHIL, an external radio frequency (RF) current with a frequency close to twice the natural frequency of the oscillator is injected. The RF signal locks the precession of the oscillators and binarizes their phase, as shown in Fig.\ref{fig:spintronic-oscillators} – the oscillator phase becomes stable at either 0 ($s_i=+1$) or 180 ($s_i=-1$) degrees relative to each other. The phase binarization induced by SHIL allows for the construction of artificial spin states in STNO and SHNO~\cite{rajabali2023injection, houshang2022prappl} arrays. By carefully designing the coupling between oscillators and controlling their phases through SHIL, it becomes feasible to find the ground state of the Ising Hamiltonian, which corresponds to the optimal solution of the mapped optimization problem. 

\subsection{Probabilistic Ising machines}

Probabilistic computing is an unconventional computing paradigm that takes the outcome of a random process and uses it to sample the possible distributions of the configuration space of a deterministic problem. The probabilistic IM (pIM) is a mapping of the Ising model into the dynamics of a stochastic network of spintronic devices~\cite{Chowdhury2023-full-stack-p-bits}. All possible steady-state configurations are sampled and a probabilistic distribution is obtained. The answer to the deterministic problem is given by the expected value of this distribution~\cite{Faria2021-hardware-Bayesian-networks}. Probabilistic computers have been put forth as purpose-built hardware accelerators that leverage the underlying physical mechanisms of spintronic oscillators to speed up computation at a lower energy cost.

\begin{figure}
    \centering
    \includegraphics[width=\linewidth]{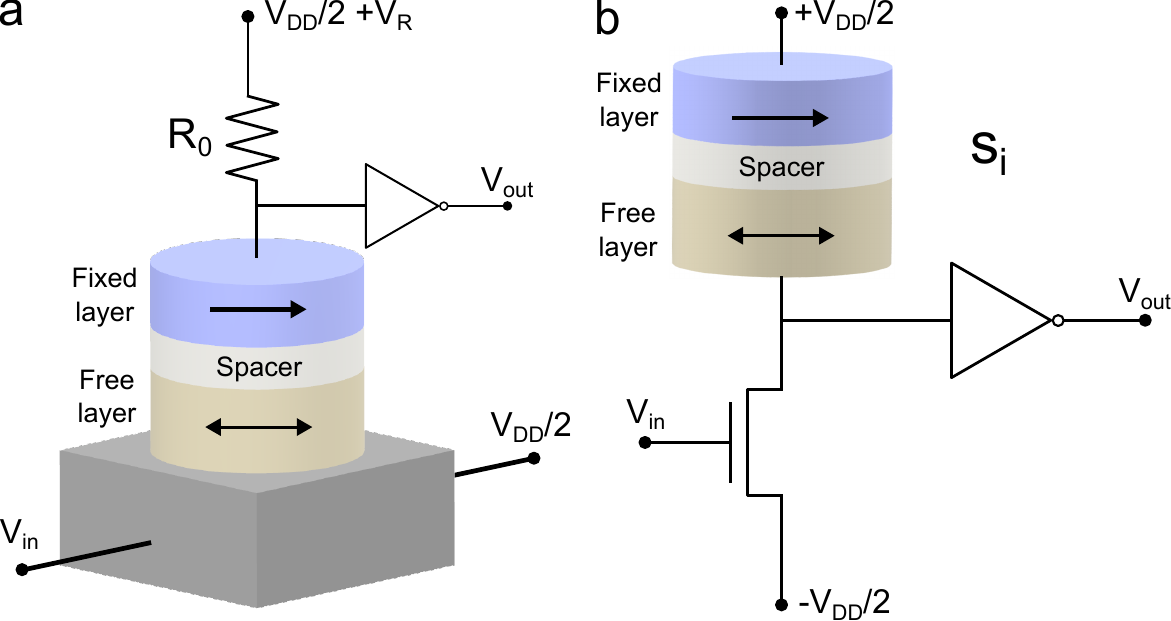}
    \caption {\textbf{MTJ-based probabilistic bits (p-bit). a.} Spin-orbit-torque-based (SOT) p-bit. Similarly to SHNOs, the heavy metal at the bottom is a source for the spin polarized current that drives the precession of the free layer. \textbf{b.} An STT-based p-bit. The precession of the free layer is driven by the spin polarized current The individual bias $h_i$ and pairwise coupling $J_{ij}$ are controlled electrically via the $V_{in}$ voltage and the operational amplifier.}
    \label{fig:p-bit}
\end{figure}

An emerging type of IM is built using arrays of electrically connected MTJs. Each MTJ in the array represents a spin in the Ising model, capable of assuming either a +1 or -1 state. These states correspond to the parallel or antiparallel alignment of the magnetizations in the two ferromagnetic layers of the MTJ, as shown in \cref{fig:p-bit}. In the context of probabilistic computing, this electrically coupled and controlled MTJ is called a probabilistic bit (p-bit), and an array such p-bits functions as a stochastic solver. For completeness, we included spin-orbit-torque (SOT) and STT driven p-bits in fig.\ref{fig:p-bit}, but they can be treated as basically the same for the purposes of building a probabilistic IM. When operating over a certain critical current, spontaneous thermally enabled magnetization reversal flips the free layer direction at random intervals, in some cases showing nanosecond random telegraph noise~\cite{Hayakawa2021-nanosecond-telegraph-noise-mtjs} in its output voltage. This voltage is then used as the input voltage of a subsequent p-bit. The strength of the coupling between p-bits is controlled by adjusting the gain of the amplifier and their electrical connections via crossbar resistors.

At room temperature, thermal fluctuations in the magnetization of the free layer introduce true randomness into the system, enabling it to explore a large number of configurations and converge towards the global minimum while escaping from local minima. This property is particularly useful for solving complex optimization problems that are typically hard for deterministic algorithms. The randomness of pIMs allows for sampling from a probability distribution defined by the energy landscape of the problem. By repeatedly initializing and letting the system relax, a wide range of solutions can be sampled, providing insights into the probability distribution of possible outcomes. This capability makes pIMs useful for tasks like optimization, sampling, and probabilistic inference, where the exploration of complex solution spaces with multiple local minima can benefit from the randomness.

\subsection{Spin-wave Ising Machines}
Spin waves (SW) and their quanta magnons are a fundamental type of collective excitations of electronic spins in magnetic systems \cite{kalinikos1986theory,chumak2017magnonic} and have been employed in the fields of signal processing \cite{papp2017SWspectrumAnalyzer,ustinov2008SWphaseShifter}, frequency synthesis \cite{litvinenko2018chaotic, litvinenko2021tunable}, logic \cite{Wang2020magnoniclogic, khitun2010magnoniclogic} and computing \cite{chumak2022advances, markovic2022SHNOneurocomp, mahmoud2020SWcomputing, watt2021SWReservoir} due to certain inherent advantages when compared to other waves. SW can be excited in a wide range of frequencies from GHz to THz and propagate with exceptionally slow velocities of kilometers to meters per second without electron transport, therefore, free from Joule heating losses making them perfect candidates for the use in delay lines and interference-based signal processing \cite{balynsky2017magnetometer, grachev2022strain, Wang2020magnoniclogic, khitun2010magnoniclogic, demokritov2003experimental, papp2017SWspectrumAnalyzer, KiechleSWOptics2023}. Moreover, SW can be efficiently controlled by electric and magnetic fields and currents~\cite{urazhdin2014nanomagnonic, EveltSWcontrol2016, vogt2014realization, demidov2015magnonic, wang2020magnonic, qin2021nanoscale} as well as strain fields acoustic waves~\cite{davies2015towards, BozhkoMagnonPhono2020, ChumakDopplerEffect2010} adding versatility to spinwave-based devices.

\begin{figure}
    \centering
    \includegraphics[width=\linewidth]{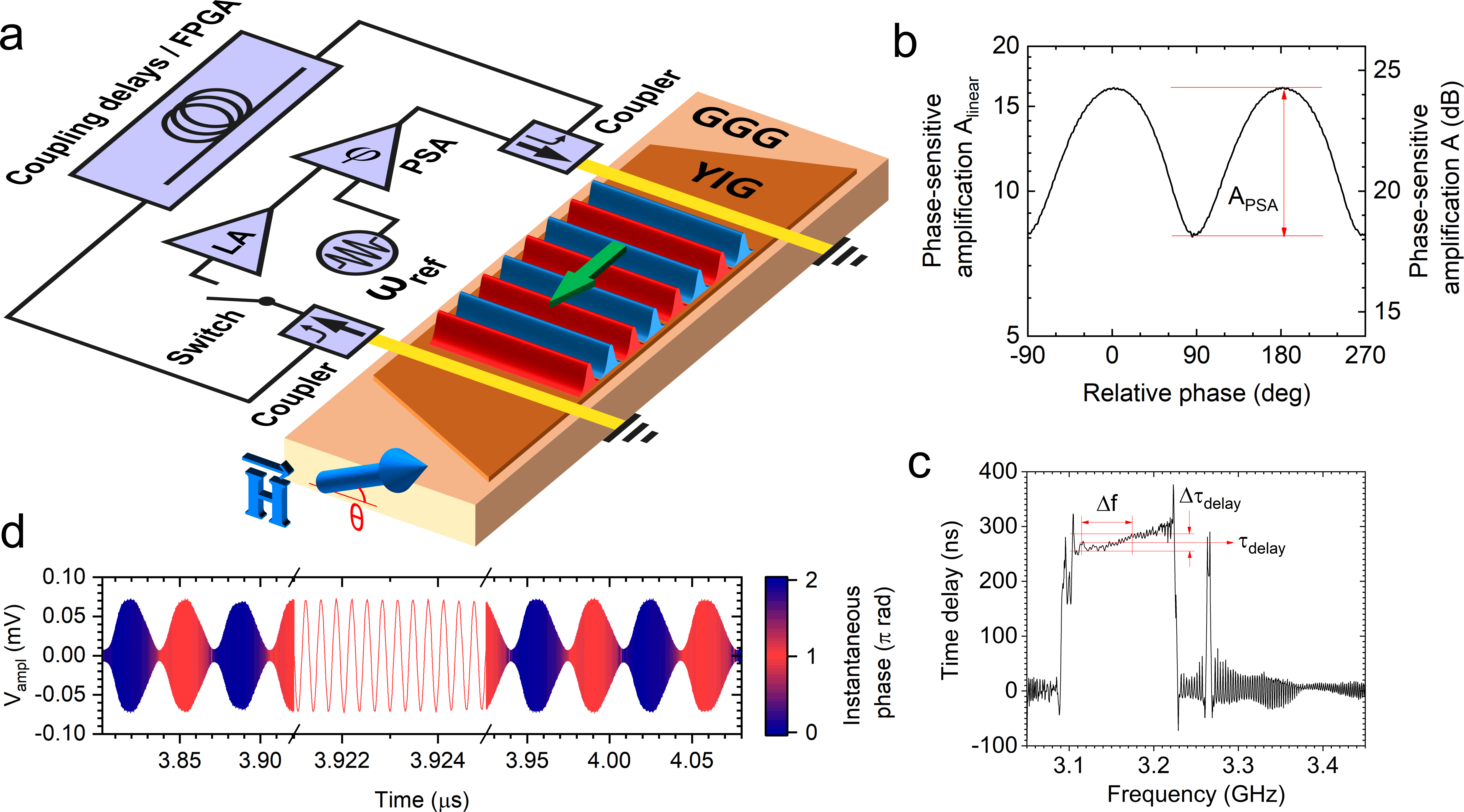}
    \caption{\textbf{A spin wave Ising machine. (SWIM) a.} Schematic of the SWIM. The artificial spins are constructed using RF spin wave (SW) pulses propagating in a YIG delay line. \textbf{b.} Phase sensitive amplification (PSA) characteristic curve. The RF pulses are binarized to be in-phase ($0^o$) or out-of-phase ($180^o$) with respect to the reference signal $\omega_{ref}$ \textbf{c.} Delay time deviation due to SW dispersion. \textbf{d.} Zoom-in of an artificial spin. The high packing efficiency of SWs in the delay line allows the SWIM to operate with much lower losses than similar setups that use laser pulses. Figures adapted from \cite{litvinenko2023spinwave} under the Creative Commons License CC BY.}
    \label{fig:SWIM}
\end{figure}

Recently, a new type of time-multiplexed IMs has been proposed~\cite{litvinenko2023spinwave, Gonzalez2023GlobalZTSWIM} utilizing dipole-dominated travelling spinwave RF pulses. With a low propagation speed ($\sim$110~km/s), these spinwave IMs are small and more thermally stable than their optical equivalent, coherent IMs (CIMs)~\cite{marandi2014networkCIM4delaylines, Honjo2021SciAdv100kCIM, Inagaki2016Sci2000CIM,McMahon2016Sci100CIM}. The schematic of the SWIM is shown in Fig.\ref{fig:SWIM}a. Each artificial spin in the spinwave-based IM (SWIM) is constructed with a RF spinwave pulse that gets phase-binarized with an external microwave phase sensitive amplifier (PSA) and propagates in an Yttrium Iron Garnet (YIG) delay line. Despite large propagation losses of SW in the YIG delay line, a conventional microwave linear amplifier (LA) can be employed to form a loop circuit with continuously circulating RF spinwave pulses; mainly due to the MHz range pulse repetition frequency and GHz range RF carrier frequency. The PSA has a two-peak amplification characteristic as a function of the phase difference between the signal of the propagating RF pulses and the external signal $\omega_{ref}$, as shown in Fig.\ref{fig:SWIM}b. The total amplification of an LA and PSA is tuned in such a way that it overcompensates for the losses of the RF pulses by 3 dB around 0$^\circ$ and 180$^\circ$ relative to the external signal $\omega_{ref}$ but leads to -3 dB round-trip losses of those signals around 90$^\circ$ and 270$^\circ$. According to the Barkhausen criterion of stability, it leads to stable generation of RF pulses having relative phase close to either 0$^\circ$ or 180$^\circ$. In the recent SWIM~\cite{litvinenko2023spinwave,Gonzalez2023GlobalZTSWIM} the interconnection between RF pulses is realized with microwave off-the-shelf components which allow forming additional coupling pulses by redirecting and delaying a portion of each propagating RF pulse with a pulse repetition delay in case of nearest-neighbour connection scheme as shown in Fig.\ref{fig:SWIM}a. We note that all-to-all interconnection is also possible with an FPGA-based measurement and feedback block as was demonstrated in~\cite{Litvinenko-50spinSAWIM}. When the SWIM gets switched on, the RF propagating pulses appear from thermal fluctuations and phase-binarize. While the initial amplitude of propagating RF pulses is small, the effective temperature of the system is high, and due to pulse interconnection, their phases actively change. Once the propagating RF pulses reach the amplitude of saturation, the signal-to-noise ratio in the system is high enough and the spin state freezes representing the solution of the mapped combinatorial problem.

It is important to highlight that the central frequency in SWIMs is 3.125 GHz while the duration of a pulse is around 20 ns (see Fig.\ref{fig:SWIM}d), which means that there are only around one hundred periods of central frequency. This is enough to consider an RF pulse to behave as a propagating time-multiplexed oscillator but at the same time low enough to cause a large phase accumulation in the system. Moreover, the duty cycle in SWIMs can be around 50\% and even less~\cite{litvinenko2023spinwave,Gonzalez2023GlobalZTSWIM} meaning that RF pulses are very closely and efficiently packed in the YIG delay line (see Fig.\ref{fig:SWIM}d). Overall, all this allows for a single trip phase accumulation of  $1.4\times10^6$ degrees. For comparison, in time-multiplexed coherent Ising machines~\cite{Honjo2021SciAdv100kCIM,marandi2014networkCIM4delaylines,Inagaki2016Sci2000CIM,McMahon2016Sci100CIM,Kako2020CIMwithErrFB,takata16bitCIMdelaylines2016,yamamoto2017CIMquantumfeedback,Haribara2016CIMPerformanceEvalDelayLines} there are around $10^3-10^4$ periods of central frequency per an artificial Ising spin with a duty cycle of 15\%~\cite{Honjo2021SciAdv100kCIM} and even 3\% in an earlier version~\cite{McMahon2016Sci100CIM}. Apart from temporal inefficiency, it leads to a very large phase accumulation of a single round trip, on the order of $10^{12} $ degrees. Despite the weak temperature expansion coefficient of optical fibers, the phase accumulation temperature coefficient in CIMs is on the order of $10^6$~degrees/K. Such a large phase accumulation temperature coefficient leads to a rapid destabilization of the loop even under small temperature fluctuations and requires the use of a phase-locked loop (PLL) stabilization system. In contrast, the phase accumulation temperature coefficient $P_{acc}TCD$ in SWIMs is only around 10~degrees/K which allows the system to operate without any PLL control under room temperature conditions.

Currently, the main limiting factor towards scalability of spinwave-based Ising machines is the nonlinear spinwave dispersion which results in a delay time deviation $\Delta\tau_{delay}$ over the operating bandwidth $\Delta f$ as shown in Fig.\ref{fig:SWIM}c. It leads to a broadening of the propagating spinwave RF pulses, which reduces their density and hence their total number in the YIG delay line. However, several papers successfully demonstrate methods that allow to compensate for the intrinsic nonlinearity and demonstrate dispersionless propagation of spinwave RF pulses~\cite{divinskiy2021dispersionless, Chen1994dispersionless}.

The recent advances in spinwave technology~\cite{wang2023perspective,grachev2024nonreciprocal,BreitbachErasing2024, grachev2023reconfigurable,grachev2022strain} pave the way to further improvement of spinwave-based time-multiplexed Ising machines. The rich physics of nonlinear spinwave phenomena~\cite{kalinikos1991envelope, ustinov2008SWsolitons, ustinov2009SWsolitons, serga2010yig} allows for the integration of phase sensitive amplification inside the YIG delay lines. Moreover, there is a prospect of using dipole-exchange and exchange-dominated spinwaves~\cite{tikhonov2020exchange, tikhonov2019spin, tikhonov2024ExchSW} with extremely slow propagation velocities and low damping~\cite{qin2022low,wang2018reconfigurable,knauer2023mKspinwave} to achieve even better miniaturization and efficiency. The recent demonstration of true amplification of spinwaves in magnonic nanowaveguides by Merbouche et al.~\cite{Merbouche2024} indicates the possibility to implement all-spinwave ring time-multiplexed oscillators without the use of inefficient transduction to electrical signals and back. Thus, SWIMs have the potential to lead in the race for miniaturized and commercially available combinatorial optimization accelerators.

\section{Network control and embeddings}
The physical platforms we have described present different challenges and opportunities based on their operational principles. A useful distinction that can be made is to classify IMs by how the artificial spin state is constructed and manipulated: it can either be represented in a physical array as a collection of spatially resolved spins or as a time-multiplexed series of RF pulsed oscillators.

\subsection{Spatially resolved Ising machines}

A spatially resolved IM is an array of spatially localized physical spin analogues that are coupled to a limited number of neighbors. As we have seen in the previous section, both p-bits and spintronic oscillators can be used to construct these networks with different coupling mechanisms and operation principles. On one hand, p-bits take advantage of the high level of maturity in the fabrication and operation of MTJs and the possibility for p-bit ensembles to operate both synchronously and asynchronously. The latter allows p-bit-based machines to massively parallelize operations and have higher throughput compared to synchronous systems. 

Oscillator arrays, on the other hand, rely on mutual synchronization and SHIL for the construction of the artificial spin state; with couplings being mediated by dipolar fields, propagating exchange waves or electrical current. Control over these couplings in STNOs and SHNOs include thermal excitation~\cite{Li2004-thermally-assisted-reversal, muralidhar2022optothermal}, voltage controlled magnetic anisotropy~\cite{Zhang2017-vcma-stnos, Fulara2020-VCMA, Gonzalez2022-VCMA-SNHO,kumar2022fabrication,choi2022voltage} and STT using memristive gates~\cite{Zahedinejad2022natmat,Khademi2023-circular-memristors, Wei2020-memristors-stnos}. Although very enticing, spintronic oscillator-based IM have only been reported based on small array measurements~\cite{houshang2022prappl} and network simulations~\cite{albertsson2021-stno-IM, mcgoldrick2022ising}, with control over their individual couplings and biases remain an engineering challenge yet to be resolved. 

Although their physical platforms are mature, spatially resolved IM present a limitation on their coupling density (i.e. on average how many pairwise connections can a spin have) due to limited connectivity since spins can only be connected to a maximum number of other spins. This limitation arises from both the array's coupling mechanisms and its distribution in space. Since the array is ordered in a lattice, there will be a limited amount of effective neighbors that any one spin can couple to. For example, an SHNO in a square lattice will have at most 8 effective neighbors since exchange and dipolar magnetic interactions drop sharply as a function of distance~\cite{Slavin2006}. Similarly, electrically connected IMs, such as p-bits, eventually run into wire routing limitations for large enough networks since the number of connections on arbitrarily coupled systems needs to grow quadratically with the number of spins. 

Since higher complexity NP-hard problems essentially require all-to-all couplings, alternative mapping procedures have been implemented to produce hardware-compatible embeddings. Sparsification is a method that addresses this problem by decreasing the number of connections per spin with the addition of ancillary spins. These problems are then embedded into the topology of the particular physical platform, for example, a Chimera or King's graph, and solved. The method of sparsification has reported very promising results for CMOS-based IMs~\cite{Aadit2022-sparsification-fpga} and integration of p-bits and sparsification has been speculated to allow the scaling of p-bit arrays into the millions of devices~\cite{Chowdhury2023-full-stack-p-bits}. However, it is worth mentioning that sparsification itself is an NP-hard problem and this adds to the complexity overhead of the operation cycle of these types of solvers~\cite{Afoakwa-BRIM}.

Likewise, the dynamic range and precision control over the coefficients in the Hamiltonian are also limitations for both physical array and physical-array-inspired IMs. Ideally, the $J_{ij}$ and $h_i$ of eq.\ref{eq:hamiltonian} are real numbers and should be encoded as such into the physical platform of choice. However, there will always be an amount of quantization uncertainty due to hardware or software limitations of actual devices and this can change the energy landscape and produce inaccurate ground state solutions. Again, with the help of ancillary spins, it is possible to reduce the bit width resolution of the IM and still have accurate solutions with the cost of exponential growth in the number of spins~\cite{Oku2020-bit-width-reduction, Yachi2022-efficient-bit-width-reduction}. The trade-off between ancillary spins and bit-width resolution, or ancillary spins and coupling density as is the case for sparsification, changes the dynamics of the system and can induce freezing of the ensemble at local minima or slow down its ground state convergence~\cite{Kikuchi2023-dynamics-bit-width-reduction}. These effects can be mitigated with the introduction of an effective temperature to change the relaxation and flipping dynamics of the individual oscillators and induce an annealing schedule. The rapid developments in these areas and physical array-based IMs are a burgeoning field of study.

\subsection{Time-multiplexed Ising machines}

The SWIM relies on a method called time multiplexing for their operation. In time-multiplexing, propagating RF pulses circulating in a loop are used to construct the artificial spin state. Phase-sensitive amplification in the ring circuit allows phase-binarization of each oscillator, turning it into an artificial Ising spin ($s_i = \pm 1$). Coupling between these individual propagating phase-binarized oscillators can be done either analogously or digitally. The analog method implies taking a small part of RF pulses, delaying them in additional delay lines, and adding them to the corresponding consequent propagating pulses via couplers. The digital method requires injecting additional RF pulses ($c_i$) to the propagating RF pulses ($s_i$) with an amplitude and phase that would correspond to the parameters of an integrated pulse resulting after the addition of all the coupling pulses coming from each propagating RF pulse according to the interconnection matrix $J_{ij}$. This can be done using digital general-purpose hardware such as matrix multiplication blocks in a field programmable gate array (FPGA), which take the input of the phase ($s_i$) of the propagating RF pulses and calculates the necessary amplitudes and phases for the feedback coupling pulses ($c_i$) proportional to $J_{ij}$ and the instantaneous spin state via:
\begin{equation}
    c_{i} =  -r\sum_{j=1}^nJ_{ij} s_j
    \label{eq:feedback-ising}
\end{equation}
\\
where the $r$ coefficient defines the strength of coupling. The design of a time-multiplexed SWIM with an FPGA measurement and feedback is shown in Fig.\ref{fig:SWIM}b. The number of maximally supported spins is defined by the delay time of the delay line (see Fig.\ref{fig:SWIM}c)and the minimum achievable RF pulse width. The time-multiplexed coupling of the spins in SWIM allows scalable network growth via FPGA-based measurement and feedback blocks. However, this scalability comes with a smaller effective interaction time between spins, inversely proportional to the number of oscillators. Thus, larger machines come at the expense of a reduced computational speed. Nevertheless, as time-multiplexed machines can support up to hundreds of thousands of spins and have the potential for miniaturization by the use of slow spin-wave, SWIMs could break the barrier of entry for commercially feasible computational accelerators for complex combinatorial problems in the foreseeable future.

An advantage of using time-multiplexing FPGA-based embeddings of the Ising Hamiltonian is that it permits dense all-to-all connectivity between spins. This allows for mappings of NP-hard problems to be embedded directly with sufficient bit-width resolution, without the need for sparsification, thus saving computational overhead during operation. The advantage of this scheme is a regular topic of discussion among current works on the subject~\cite{Aadit2022-sparsification-fpga, grimaldi2023evaluating, Chowdhury2023-full-stack-p-bits, mohseni2022ising}. In the next section, we contribute to these discussions by benchmarking the schemes we have presented so far and how their prospects fare for further developments within the space of hardware computing accelerators.

\section{Benchmarking}
In Table 1, we benchmark spintronic-based schemes and compare them with other technologies. Here we focus on the following system parameters: operational speed (including clock frequency and time-to-solution), energy efficiency (system power consumption, energy required to reach solution, energy efficiency), system size, potential for scalability (including number of Ising spins and connectivity), and system stability. We have divided the list of spintronic IMs into two groups - spatial physical array systems and time-multiplexed IMs. 

Ising machines implemented as physical oscillator arrays allow for the fastest time-to-solution parameter as compared to any other technologies. Two main factors contribute to this. First, it is their operation at high GHz frequency and very fast frequency tunability~\cite{litvinenko2020ultrafast,litvinenko2022ultrafast} and response which allows them to change their phase state on the nanosecond time scale. Second, IMs implemented as physical oscillators or p-bits can continuously interact with each other in contrast to time-multiplexed IMs. Continuous interaction comes at the price of interconnection complexity with subsequent limitations for the number of supported Ising spins, which is especially applicable for GHz frequency oscillators where the different lengths of connection may introduce considerable parasitic time delays and unintended couplings.

On the other hand, time-multiplexed IMs can be scaled up to tens and hundreds of thousands of artificial spins but, as it was explained in the previous section, have the disadvantage of reduced interaction time that strongly affects their time-to-solution.

pIMs have parameters that are in between physical oscillator arrays and time-multiplexed spinwave oscillator-based IMs primarily due to their relatively slow switching speed and the probabilistic principle of operation, with noise-induced flips slowing down convergence to the optimal solution. It allows for less problematic interconnection and better scalability but dramatically increases time-to-solution. Nevertheless, low power consumption, verified CMOS integration capabilities, and robustness of their operational principle makes MTJ p-bits a strong competitor in the world of IMs.

Overall, the main advantages of spintronic components over conventional CMOS and optic technologies are miniaturization, high operational frequency for spintronic oscillators, and relatively low power consumption for MTJ p-bits. It allows spintronic components to lead in each category - spatial and time-multiplexed architectures of IMs. 

\begin{figure}
    \centering
    \includegraphics[width=0.9\linewidth]{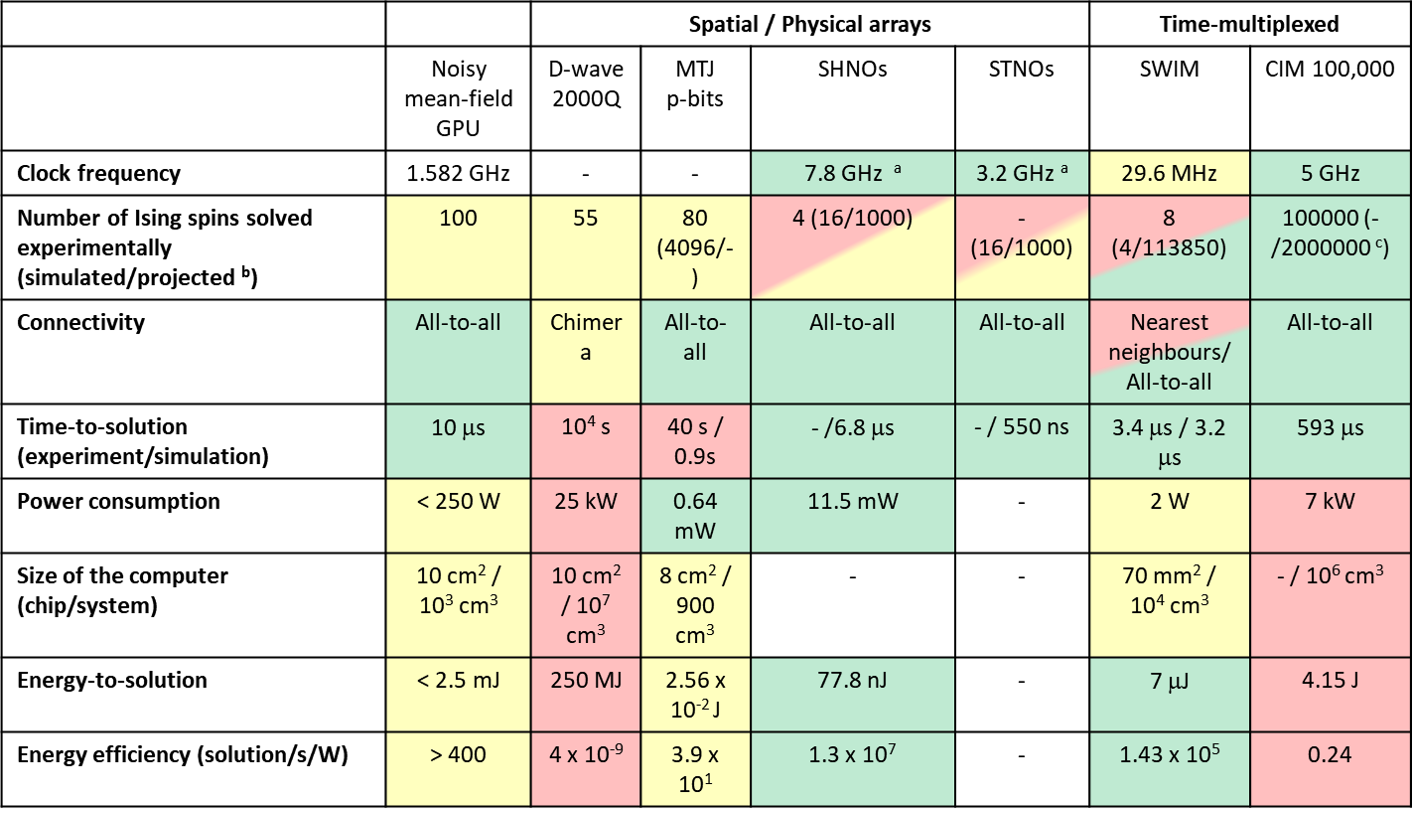}
    \caption*{Table 1: Bench-marking of spintronic-based Ising machines. 
$^{a}$ - Oscillation frequency of phase-transition oscillators.
$^{b}$ - Projected number of artificial spins according to the potential in scalability }
    \label{table:BM}
\end{figure}

\section{Outlook} 
As we have presented in the preceding sections, the interest in spintronic oscillators for combinatorial optimization has been steadily increasing over the past decade given both advances in these technologies and their high compatibility with existing CMOS devices. This compatibility is quite important for further developments and hybrid architectures since Ising machines can be programmed to explore vast configuration spaces via effective embeddings, but are less effective than conventional electronics in deterministic tasks such as Boolean logic and number addition. This difference lies in the fact that randomness and computation load are highly parallelized in the Ising machine concept and as such it presents an opportunity with a currently relevant and trending topic: machine learning.

A Hopfield network (HN)~\cite{Hopfield1982-hopfield-networks} is a type of associative all-to-all connected neural network that was among the earliest developments in modern machine learning. Also called an \textit{Ising neural network}, the HN was indeed inspired by the idea of the Ising Hamiltonian of eq.\ref{eq:hamiltonian} and applied it to biologically inspired learning by replacing the spins by neurons, the couplings by weights and the biases by thresholds. One cycle of the HN is equivalent to the relaxation process of an IM, with the key difference of the HN being recursive, with its output being fed into the input and iterated again. This is because the cost function of the HN is not the Ising Hamiltonian but instead the weights of the system are changed after each iteration of minimizing an arbitrary function. In this sense, a HN can be seen as a reprogrammable IM with an external memory element. Therefore, spintronic hardware developed for IM can also be used for neural networks and machine learning algorithms. It has been shown that IMs can be adapted to run machine learning algorithms~\cite{Laydevant2023-ising-machine-equilibrium-propagation} and their applicability to cutting-edge algorithms is a fertile line of research for the future.

Although size and coupling density are frequently reported metrics in IM, exploration of applied mathematical concepts such as higher-order coupling and multivalued spins have been predicted to enable the construction of novel IM architectures that can perform complex optimization tasks in smaller systems. Higher-order coupling refers to the possibility of constructing a Hamiltonian with terms that contain multiplication of more than two spins. For example, the term $\sum J_{ijk} s_i s_j s_k$ has cubic coupling encoded in tensor $J_{ijk}$. Multivalued spins refer to spins beyond binary, for example, $s_i \in -[1,0,1]$, and could used to construct trinary-state IMs. Although just theorized for now, exploration of these two concepts can help demonstrate the intrinsic value of using spintronic devices for computation since there are many opportunities in using all-analog accelerators. 

Due to the high compatibility mentioned above, many of the proofs-of-concept mentioned before employ existing reconfigurable chips such as microcontrollers or FPGAs to perform updates or to couple the system. Although this has produced many scientifically sound and interesting advances, for the field of spintronic acceleration to fully explore the computational speed up and parallelization that should come from the physical embedding of Ising models, less dependence on digitization should be encouraged. As it stands, the same bottlenecks of sequential processing and digitation quantization uncertainty that apply to conventional computing accelerators apply to IM. More research is needed in the field of all analog magnetic computation, with time evolution and convergence of the spin array arising from magnetic dynamics. A source of inspiration could be the field of photonics-based Ising machines. Analog computation of state evolution has permitted implementations of space-multiplexed IMs with a high potential for spin scalability, at the cost of actual device size. Advances in magnonic waveguides, resonators, and more complicated components are indicators that spintronics and magnonics are reaching a level of technological maturity that can be channeled toward unconventional computing schemes. 

\section{Conclusion}
As we have shown, computation acceleration using spintronic devices is an active and growing field of research in spintronics and magnetic materials. Due to operational and manufacturing constraints, alternatives to conventional computing are being explored for high throughput applications, such as machine learning, magnetics-based in-memory computing is an appealing candidate for realizing low footprint and scalable IMs. Further innovations in the underlying physical mechanisms of the  platforms that can host IMs, for example, material fabrication or manipulation of magnetic phenomena, can improve the effectiveness of the spintronics-based IM.

Although spintronic oscillators and magnonic delay lines are broad subjects of study by the academic and industrial communities and their advances continue to show promise as complements or alternatives to existing CMOS technology, they have yet to demonstrate an application that they are uniquely suited for and would ramp up their development. In our opinion, a spintronic-based IM is that application and can act as a low-cost room-temperature noise-resistant alternative to the various quantum annealers. Given that the periodic claims of quantum supremacy are highly subjective and are routinely shown to be matched by medium-sized computers or even laptops, we think that the implementation of magnetic materials and concepts can enable the realization of large-scale devices that can tackle truly massive problems and show non-subjective improvements to conventional computer architectures. 

The flexibility of spintronic and magnonic hardware allows for the construction of spatially localized and time-multiplexed computational schemes and advancements in the physical platforms hosting these embeddings can also help in establishing standard benchmarks for comparison of the two. It is likely that some problems are more suited for embedding in one architecture or the other and knowing their strengths can help tailor the research to real-world use cases. Furthermore, due to this flexibility, novel ideas such as high-order and multistate IMs can be realized and studied, bringing much-needed variety to the slowly plateauing silicon-dominated field of von Neumann computation. Future general-purpose computers may be hybrid, with some purpose-built spintronic accelerators controlled by a central processor. The speed-ups and power savings afforded by massive parallelization are apparent for high-performance computing, but unconventional architectures also have the potential to revolutionize research-driven applications and even personal computing. There is much to advance towards a future where magnetism and magnetic materials can be seamlessly incorporated into the next-generation computation accelerators.

\section*{Acknowledgments}
This work was partially supported by the Horizon 2020 research and innovation program No.~835068 “TOPSPIN” and the ERC Proof of Concept Grant No.~101069424 "SPINTOP" and the Marie Skłodowska-Curie grant agreement No.~101111429 "SWIM". This work was also partially supported by the Swedish Research Council (VR Grant No. 2016-05980).

\section*{Conflicts of interest}
A.L., R.K. and J.Å are inventors of a Swedish patent application (FIK-014SE/2250979-8) that covers the implementation of a
spinwave Ising machine using propagating spinwave RF pulses. A.L., R.K. and J.Å. are also co-founders of a spin-off company that aims to develop spinwave-based Ising Machines. All other authors declare no competing interests.

\bibliographystyle{elsarticle-num} 
\bibliography{cas-refs}





\end{document}